\documentclass[conference]{IEEEtran}
\IEEEoverridecommandlockouts
\usepackage{cite}
\usepackage{amsmath,amssymb,amsfonts}
\usepackage{algorithmic}
\usepackage{graphicx}
\usepackage{textcomp}
\usepackage{xcolor}
\usepackage{subfigure}
\usepackage{booktabs}
\usepackage[colorlinks=true,linkcolor=black,anchorcolor=black,citecolor=black]{hyperref}
\usepackage{multirow}
\usepackage{tablefootnote}
\usepackage [misc] {ifsym}

\def\BibTeX{{\rm B\kern-.05em{\sc i\kern-.025em b}\kern-.08em
  T\kern-.1667em\lower.7ex\hbox{E}\kern-.125emX}}
\begin{document}

\title{Heterogeneous Directed Hypergraph Neural Network over abstract syntax tree (AST) for Code Classification\\
\thanks{DOI reference number: 10.18293/SEKE2023-136}
}

\author{\IEEEauthorblockN{Guang Yang\IEEEauthorrefmark{1}, Tiancheng  Jin\IEEEauthorrefmark{1}, Liang Dou\IEEEauthorrefmark{1}\IEEEauthorrefmark{2}$^{(\textrm{\Letter})}$
}
\IEEEauthorblockA{\IEEEauthorrefmark{1} School of Computer Science and Technology, East China Normal University, Shanghai, China}
\IEEEauthorblockA{\IEEEauthorrefmark{2} NPPA Key Laboratory of Publishing Integration Development, ECNUP, Shanghai, China}
\IEEEauthorblockA{51215901104@stu.ecnu.edu.cn,  51184506019@stu.ecnu.edu.cn, ldou@cs.ecnu.edu.cn}
}

\maketitle

\begin{abstract}
Code classification is a difficult issue in program understanding and automatic coding. Due to the elusive syntax and complicated semantics in programs, most existing studies use techniques based on abstract syntax tree (AST) and graph neural network (GNN) to create code representations for code classification. These techniques utilize the structure and semantic information of the code, but they only take into account pairwise associations and neglect the high-order data correlations that already exist between nodes of the same field or called attribute in the AST, which may result in the loss of code structural information. On the other hand, while a general hypergraph can encode high-order data correlations, it is homogeneous and undirected which will result in a lack of semantic and structural information such as node types, edge types, and directions between child nodes and parent nodes when modeling AST. In this study, we propose a heterogeneous directed hypergraph (HDHG) to represent AST and a heterogeneous directed hypergraph neural network (HDHGN) to process the graph for code classification. Our method improves code understanding and can represent high-order data correlations beyond paired interactions. We assess our heterogeneous directed hypergraph neural network (HDHGN) on public datasets of Python and Java programs. Our method outperforms previous AST-based and GNN-based methods, which demonstrates the capability of our model.
\end{abstract}

\begin{IEEEkeywords}
hypergraph, heterogeneous graph, code classification, graph neural networks, code representation
\end{IEEEkeywords}

\section{Introduction}
With the advancement of modern computer software, how to learn from vast open-source code repositories to enhance software development has become an essential research topic. In recent years, source code processing, which tries to help computers automatically comprehend and analyze source code, has received a lot of attention. Several works have been suggested including code classification \cite{Mou2016ConvolutionalNN,Gilda2017SourceCC,Vagavolu2021AMO,Wang2022LearningTR,Zhang2019ANN,Lu2021StudentPC}, method name prediction \cite{Vagavolu2021AMO}\cite{Wang2022LearningTR}\cite{alon2018codeseq}\cite{Alon2019code2vecLD}, code summarization \cite{Vagavolu2021AMO}\cite{Hu2018DeepCC}\cite{Liu2021RetrievalAugmentedGF} and code clone detection \cite{ Zhang2019ANN}\cite{Jiang2007DECKARDSA}\cite{White2016DeepLC}, etc.

Due to the improvement of machine learning technology, particularly deep learning, more and more work has employed deep learning for code classification. Currently, there are two main categories of code classification methods: AST-based and GNN-based. To take advantage of the semantic and structural information of the source code, several studies adopt AST when learning code representations \cite{Mou2016ConvolutionalNN}\cite{Zhang2019ANN}\cite{alon2018codeseq}\cite{Alon2019code2vecLD}. Some research uses graph neural network (GNN) to create code representations for code categorization to a better understanding of the structure of code based on AST \cite{Vagavolu2021AMO}\cite{Wang2022LearningTR}\cite{allamanis18learning}\cite{Long2022MultiViewGR}.

\begin{figure}[t]
\centerline{\includegraphics[scale=0.4]{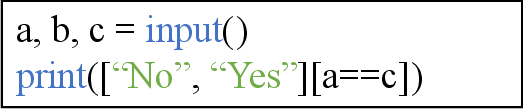}}
\caption{An example of the code snippet. The program reads three inputs a, b, and c in turn, if a equals c, ``Yes" will be output, otherwise, ``No" will be output. }
\label{fig1}
\end{figure}

\begin{figure*}[t]
\subfigure[The AST created by offical python module.]{
\begin{minipage}[t]{0.24\linewidth}
\centering
\includegraphics[scale=0.4]{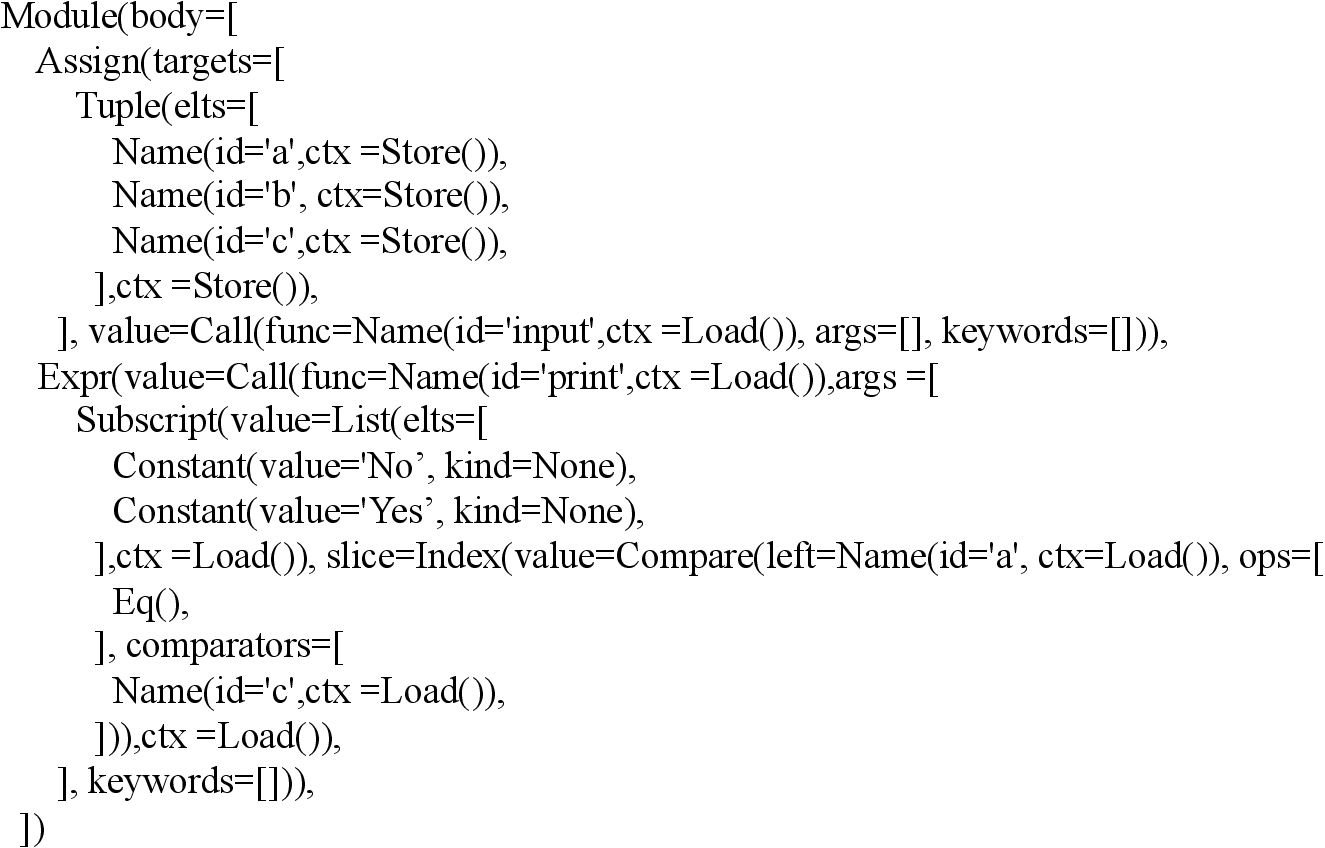}
\end{minipage}
}
\quad\quad\quad\quad\quad\quad\quad\quad\quad\quad\quad\quad\quad
\subfigure[The illustration of the AST.]{
\begin{minipage}[t]{0.25\linewidth}
\centering
\includegraphics[scale=0.24]{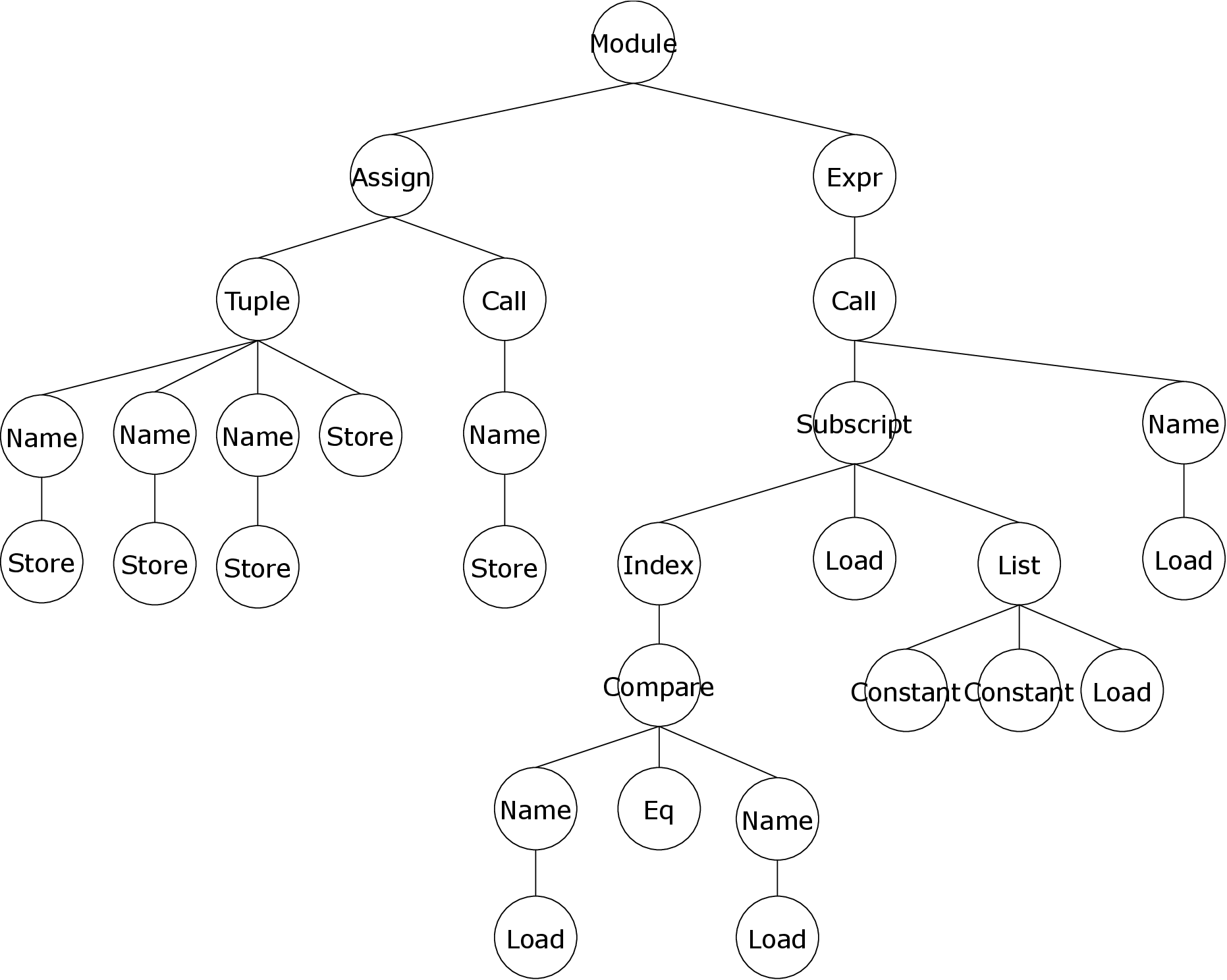}
\end{minipage}}
\caption{The AST of code snippet in Fig.~\ref{fig1}. We use the official python module to print the AST to depict the details in Fig.~\ref{fig2}(a). We draw the illustration of the AST in Fig.~\ref{fig2}(b) to demonstrate the parent-child relationship between AST nodes.}
\label{fig2}
\end{figure*}

\begin{figure}[b]
\centerline{
\subfigure[The pairwise relationships.]{\includegraphics[scale=0.31]{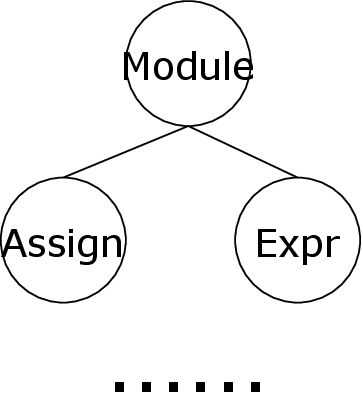}}
\quad\quad\quad
\subfigure[The high-order data correlation.]{\includegraphics[scale=0.31]{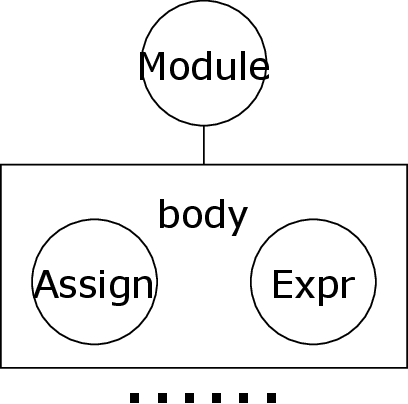}}}
\caption{The pairwise connections and high-order data correlation between three AST nodes. We ignore other AST nodes in the figure.}
\label{fig3}
\end{figure}

Although these AST-based and GNN-based techniques employ the structural information of source code and demonstrate their effectiveness, there is a problem that they only take into the pairwise relationships and ignore the possible high-order correlations between AST nodes of the same field or called attribute. When code is parsed into an AST, each parent AST node has child AST nodes belonging to various fields. A parent node may have several child nodes under the same field, and these nodes have high-order correlations with one another. Fig.~\ref{fig1} depicts an example of a python code snippet. The corresponding AST generated by the official python ast module\footnote{https://docs.python.org/3/library/ast.html.} is illustrated in Fig.~\ref{fig2}(a). As we can see, the ``Module" is the root node of the AST. It has two child nodes, ``Assign" and ``Expr" which belong to the field named ``body." When modeling the correlations between the three nodes, previous approaches only consider the pairwise relationships, i.e., the pair of ``Module" and ``Assign" and the pair of ``Module" and ``Expr," as demonstrated in Fig.~\ref{fig3}(a). The high-order correlation that ``Assign" and ``Expr" both belong to the ``body" of ``Module" as shown in Fig.~\ref{fig3}(b) is dismissed. This may result in the loss of code structural information.

\begin{figure}[t]
\centerline{
\subfigure[The general hypergraph.]{\includegraphics[scale=0.3]{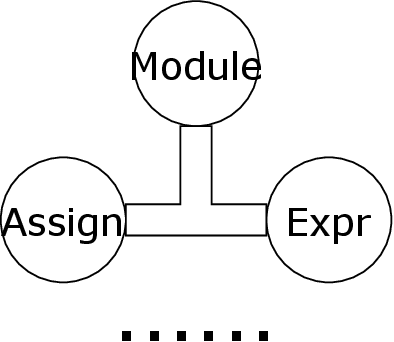}}
\quad\quad\quad
\subfigure[The HDHG.]{\includegraphics[scale=0.3]{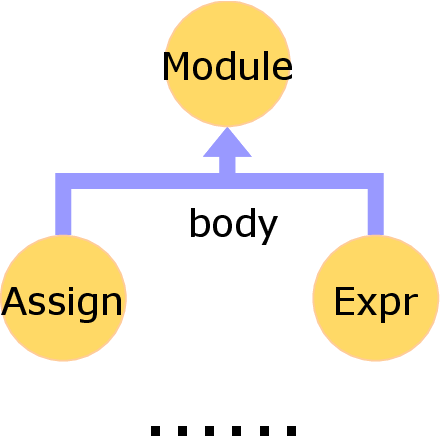}}}
\caption{Comparison of using general hypergraph and HDHG to model the relationships between three nodes of Fig.~\ref{fig3}. The difference is that hyperedges in HDHG have direction and type.}
\label{fig4}
\end{figure}

In recent years, hypergraph, which can encode high-order data correlations, has drawn a lot of interest. Considering the outstanding performance of hypergraph in graph classification \cite{Huang2021UniGNNAU}\cite{Feng2018HypergraphNN}, we present hypergraph into code classification. On the other hand, a general hypergraph is homogeneous and undirected, i.e., it only has one type of node and one type of edge, and its hyperedge is undirected. If we represent the AST with a general hypergraph, it will result in lack of semantic and structural information such as field names and directions. As illustrated in Fig.~\ref{fig4}(a), a typical hypergraph does not have the field name and the direction to show who is the parent node and who is the child node.

To tackle above problems, we propose a heterogeneous directed hypergraph (HDHG) to model the AST and an HDHGN for code classification. First, we propose to use heterogeneous directed hyperedge to show the relationship of AST nodes, the example of Fig.~\ref{fig3} is shown in Fig.~\ref{fig4}(b). Second, we combine deep learning techniques from hypergraph neural networks and heterogeneous graph neural networks to create the HDHGN, and we also add operations to process directed hyperedge.

We evaluate our method on public datasets Python800 and Java250 \cite{puri2021codenet}. Our model gets 97\% in accuracy on Python800 and 96\% on Java250 which outperforms previous state-of-the-art (SOTA) AST-based and GNN-based work. Our study demonstrates the utility of HDHG and HDHGN for code classification.

The main contributions of this paper are:
\begin{itemize}
\item We propose an HDHG to depict AST.
\item We propose an HDHGN to generate vector representations for code classification.
\item We assess our model on public datasets and compare it with previous state-of-the-art (SOTA) AST-based and GNN-based methods.
\end{itemize}

\section{Related Work}
Code classification is to classify codes based on their functions. Different from natural language, code has structural information. As a result, several works adopt AST by various techniques. 
Mou et al. \cite{Mou2016ConvolutionalNN} is one of the first works to suggest a tree-based convolutional neural network (TBCNN) in code classification. Alon et al. propose code2seq \cite{alon2018codeseq} and code2vec \cite{Alon2019code2vecLD} to deconstruct code to a collection of paths in its AST. J. Zhang et al. \cite{Zhang2019ANN} propose a novel neural network called ASTNN for source code representation for code classification and clone detection. N. D. Q. Bui et al. \cite{Bui2021TreeCapsTC} propose TreeCaps by fusing capsule networks with tree-based convolutional neural network (TBCNN) in code classification.

With the popularity of GNN, more works apply kinds of GNN in code classification 
based on AST to strengthen the comprehension of code structures. M. Allamanis et al. \cite{allamanis18learning} first construct graphs from source code by adding edges like control flow and data flow to AST and employing a gated graph neural network (GGNN) to process program graphs. V. Hellendoorn et al. \cite{Hellendoorn2020GlobalRM} propose a model called GREAT based on the transformer architecture by extracting global relational information from code graphs. D. Vagavolu et al. \cite{Vagavolu2021AMO} propose an approach that can extract and use program features from multiple code graphs. M. Lu et al. \cite{Lu2021StudentPC} improve gated graph neural network (GGNN) in program classification. T. Long et al. \cite{Long2022MultiViewGR} propose a multi-view graph program representation method that combines both data flow and control flow as multiple views and apply GNN to process. W. Wang et al. \cite{Wang2022LearningTR} propose to leverage heterogeneous graphs to show code based on previous work and adopt heterogeneous GNN to process.

\begin{figure*}[t]
\centerline{\includegraphics[scale=0.6]{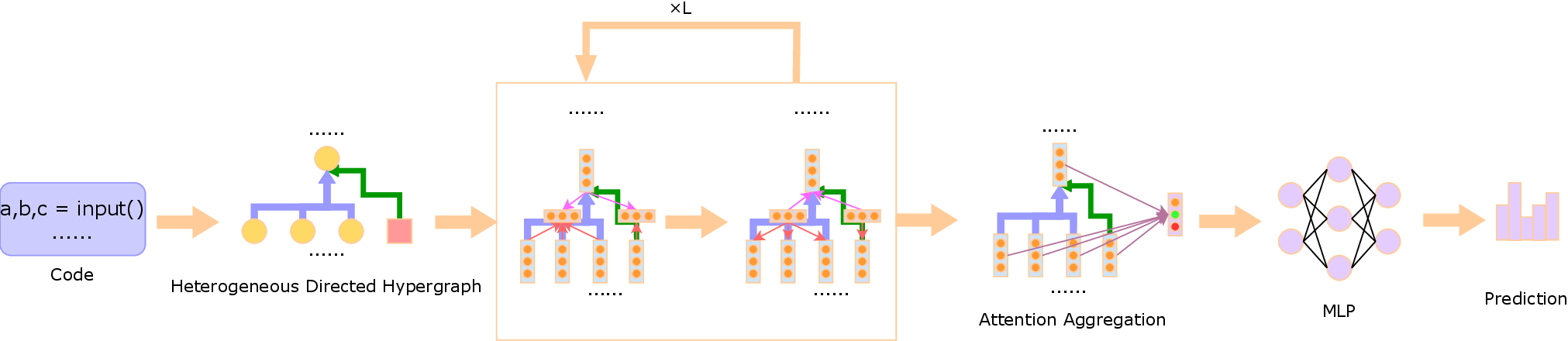}}
\caption{Overview of the process.}
\label{fig5}
\end{figure*}

\section{Preliminary}
In this section, we introduce some fundamental background ideas, such as hypergraph, heterogeneous graph, and AST.

\subsection{Hypergraph}
In an ordinary graph, an edge can only be connected with two vertices. Different from general graphs, the edge of a hypergraph \cite{Berge1973GraphsAH} can link any number of vertices. Formally, a hypergraph $H$ is a pair $H = (V,E)$ where $V$ is a set of elements called nodes or vertices, and $E$ is a set of non-empty subsets of $V$ called hyperedges or links.

A directed hypergraph \cite{Gallo1993DirectedHA} is a hypergraph with directed hyperedges. A directed hyperedge or hyperarc is an ordered pair, $E = (X,Y)$, of (possibly empty) disjoint subsets of vertices; $X$ is the tail of $E$ while $Y$ is its head. A backward hyperarc, or simply B-arc, is a hyperarc $E = (X,Y)$ with $|Y| = 1$. A forward hyperarc, or simply F-arc, is a hyperarc $E = (X,Y)$ with $|X| = 1$. A hypergraph whose hyperarcs are B-arcs is known as a B-graph (or B-hypergraph). A hypergraph whose hyperarcs are F-arcs is known as an F-graph or F-hypergraph.

In our study, since the child node of AST points to the parent node and the child node has only one parent node, our HDHG is a B-hypergraph.

\subsection{Heterogeneous Graph}
A heterogeneous graph \cite{Sun2012MiningHI} is a graph consisting of multiple types of entities or nodes and multiple types of links or edges. A heterogeneous graph is represented as $G = (V, E)$, consisting of an entity set $V$ and a link
set $E$. A heterogeneous graph is also correlated with a node type mapping function $\phi : V \to A$ and a link type mapping function $\psi : E \to R$. $A$ and $R$ represent the sets of predefined object types and link types, where $|A|+|R| > 2$.

\subsection{Abstract Syntax Tree}
The AST represents the source code's abstract syntax structure. The code compiler will parse the code into an AST through the program syntax and semantic analysis. Each node on the tree represents a structure in the source code and belongs to different AST node types. Each AST node has zero, one, or several fields that can be thought of as the node's attributes. Each field may have none, one, or a list of objects such as AST node, number, and string. If one AST node contains a field with a different AST node, and the latter is equivalent to the former's child AST node.

\section{Methodology}\label{Methodology}
We first convert the code snippet into an AST and construct an HDHG based on it, then put it into our HDHGN. We aggregate the vector representations for code categorization once we get the network's node's vector representation. The overview of our model is demonstrated in Fig.~\ref{fig5}.

\subsection{Heterogeneous Directed Hypergraph}\label{Heterogeneous Directed Hypergraph}
\begin{figure}[t]
\centerline{\includegraphics[scale=0.23]{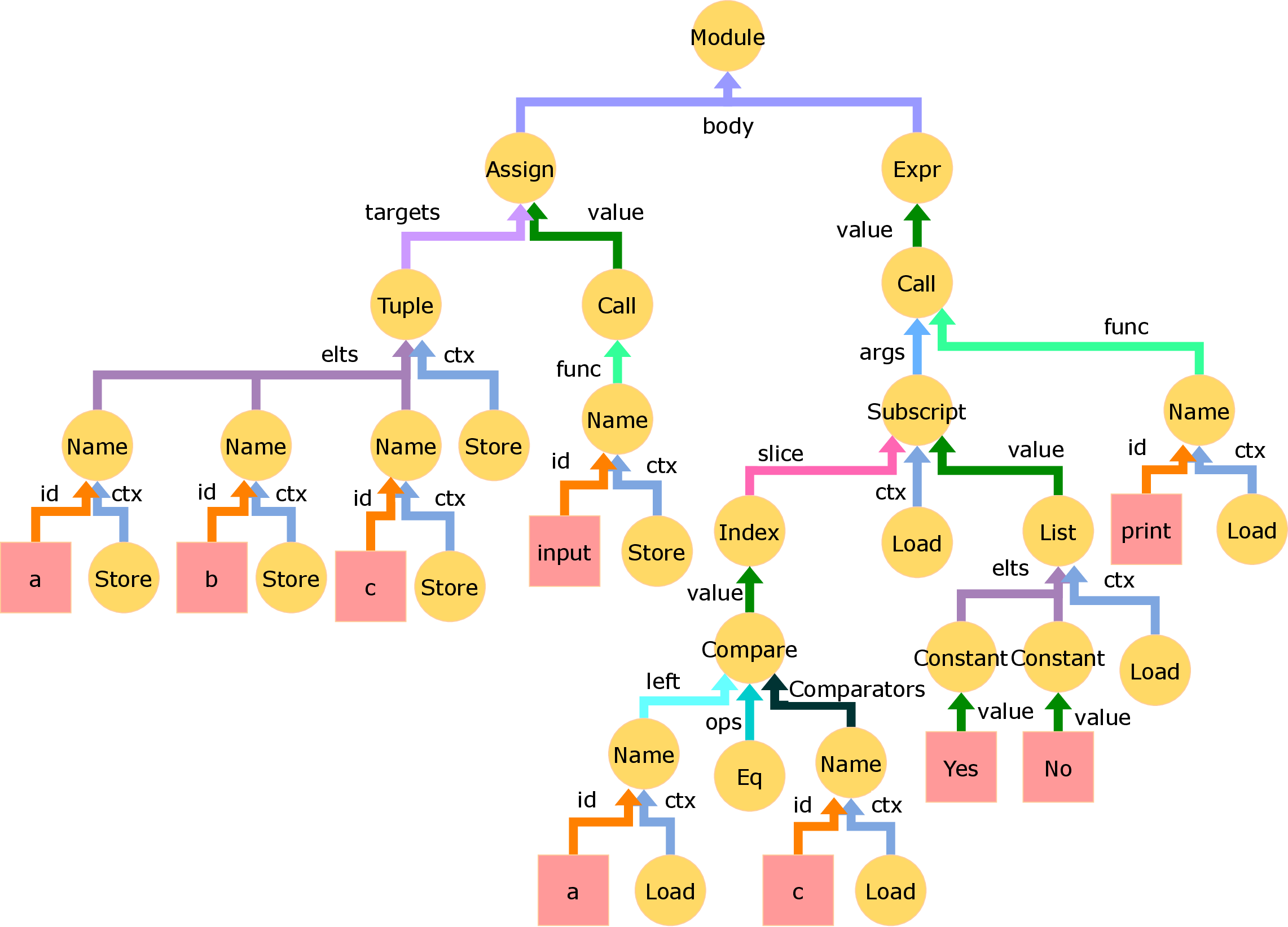}}
\caption{The HDHG of AST in Fig.~\ref{fig2}. The circular node is the ``AST'' node, and the square node is the ``identifier'' node. The edge can connect multiple nodes, the node connected with an arrow is the head node of the edge, and the node connected without the arrow is the tail node. Various hues correspond to various edge kinds. The node and edge both display the node value and edge type.}
\label{fig6}
\end{figure}

We parse the code snippet into an AST with a code compiler, then we develop the HDHG based on the AST. We set the node of AST as the ``AST" node and the identifier of AST as the ``identifier" node in HDHG. We set the value of the ``AST" node as its AST node type name, set the value of the ``identifier" node as its content, and treat them as two different types of nodes. The field is configured as a directed hyperedge. If one node has a field including another node, the latter node belongs to the tail of the field hyperedge, the former is the head of the field hyperedge. We designated the field name as the type of hyperedge. The illustration of the HDHG of AST in Fig.~\ref{fig2} is shown in Fig.~\ref{fig6}.

\subsection{Heterogeneous Directed Hypergraph Neural Network}\label{Heterogeneous Directed Hypergraph Neural Network}
\subsubsection{Definition}
We let an HDHG $G = (N, E)$, which includes a node set $N = \left\{ n_{1},~n_{2},\ldots,n_{|N|} \right\}$ and a directed hyperedge set $E = \left\{ e_{1},e_{2},\ldots,e_{|E|} \right\}$. Each node $n = (\mu,x)$, where $\mu$ represents node type and $x$ is the value of the node. Each directed hyperedge $e = \left( {\rho,S(e),T(e)} \right)$, $\rho$ represents edge type, $S(e) = \left\{ n_{1},\ldots,n_{|{S(e)}|} \right\} \subseteq N$ is the tail nodes of hyperedge $e$, $T(e) \in N$ is the head node of hyperedge $e$, they show the direction of the hyperedge $e$ is $S(e)$ to $T(e)$.

\subsubsection{Feature Initialization}
According to the value $x$ and the category $\mu$ of node $n$, we obtain embedding vector $d_{n} \in \mathbb{R}^{C_{1}}$ by embedding function as \eqref{eq1}, where $C_{1}$ is the dimension size of the embedding vector.
\begin{equation}
d_{n} = {Embed}_{\mu}(x)\label{eq1}
\end{equation}
To put embedding vectors of various types into the same vector space, we make a linear projection to obtain the initial feature vector $h^{0}_{n} \in \mathbb{R}^{C_{2}}$ of node $n$ based on the corresponding node type $\mu$ as \eqref{eq2}, where $C_{2}$ is the dimension size of feature vector and hidden vector.
\begin{equation}
h_{n}^{0} = W_{\mu}d_{n} + b_{\mu}\label{eq2}
\end{equation}
We also obtained embedding vector $d_{e} \in \mathbb{R}^{C_{2}}$ of hyperedge $e$ according to the edge type $\rho$ as \eqref{eq3}.
\begin{equation}
d_{e} = {Embed}_{edge}(\rho)\label{eq3}
\end{equation}

\subsubsection{Heterogeneous Directed Hypergraph Convolution Layer}
Our model updates each node vector in heterogeneous directed hypergraph convolution (HDHGConv) layers. We refer to the framework of two-stage message passing of hypergraph neural network \cite{Huang2021UniGNNAU} which has two steps: aggregating messages from nodes to hyperedges and aggregating messages from hyperedges to nodes. Differently, we add operations that add heterogeneous information and direction information.

\textbf{Aggregating messages from nodes to hyperedges:} First, the hidden vector $h_n^{l-1}$ of each node $n$ is multiplied by the head matrix or tail matrix to get the message vector $m^l_{n\_e}$ from node $n$ to hyperedge $e$ as \eqref{eq20}, where $l = 1,~2,\ldots,L$ indicate layer number, $L$ is the total number of layers.
\begin{equation}
m_{n\_e}^{l} = \left\{
\begin{aligned}
W_{head}^{l}h_{n}^{l-1} + b_{head}^l\, if \ n \in S(e)\\
W_{tail}^{l}h_{n}^{l-1} + b_{tail}^l\, if \ n=T(e)
\end{aligned}
\right.
\label{eq20}
\end{equation}
Directed hyperedge $e$ gathers messages from its tail nodes and head node by transformer attention mechanism \cite{Vaswani2017AAN}. For each message, the attention score is formulated as \eqref{eq4}, where $d_{e}$ is the edge type vector.
\begin{equation}
\begin{aligned}
\alpha_{n\_e}^{l} &= Softmax\left( \frac{{\left( W_{q1}^{l}d_{e}\right)}^{T}W_{k1}^{l}m_{n\_e}^{l}}{\sqrt{C_{2}}} \right)
\label{eq4}
\end{aligned}
\end{equation}
We obtain the vector $o_e^l$ of directed hyperedge $e$ as \eqref{eq6}.
\begin{equation}
o_{e}^{l} = {\sum_{n \in S(e)\ or\ n=T(e)}{\alpha_{n\_e}^{l}W_{v1}^{l}m_{n\_e}^{l}}}\label{eq6}
\end{equation}
Then we add edge type vector to $o_e^l$ as \eqref{eq16}, where $z_e^l$ is formed as \eqref{eq7}.
\begin{equation}
q_{e}^{l} = o_{e}^{l}+z_e^l\label{eq16}
\end{equation}
\begin{equation}
z_e^l = W_z^{l}d_{e}+b_{z}^l\label{eq7}
\end{equation}

\textbf{Aggregating messages from hyperedges to nodes: }For each directed hyperedge $e$ whose head node or tail node is $n$, the $q_j^l$ will be linear projected by \eqref{eq17} to get message $m_{e\_n}^l$ which will be sent to $n$.
\begin{equation}
m_{e\_n}^{l} = \left\{
\begin{aligned}
W_{to\_head}^{l}q_{e}^{l} + b_{to\_head}^l\, if \ n \in S(e)\\
W_{to\_tail}^{l}q_{e}^{l} + b_{to\_tail}^l\, if \ n=T(e)
\end{aligned}
\right.
\label{eq17}
\end{equation}
Same as before, we aggregate messages to get $v_n^l$ by the transformer attention mechanism.
\begin{equation}
\begin{aligned}
\alpha_{e\_n}^{l} &= Softmax\left( \frac{{\left( W_{q2}^{l}h_{n}^{l-1}\right)}^{T}W_{k2}^{l}m_{e\_n}^{l}}{\sqrt{C_{2}}} \right)
\\
\label{eq10}
\end{aligned}
\end{equation}
\begin{equation}
v_{n}^{l} = {\sum_{T(e) = n \ or \ n \in S(e)}{\alpha_{e\_n}^{l}W_{v2}^{l}m_{e\_n}^{l}}}\label{eq11}
\end{equation}

Last, we update hidden vector $h_n^l$ of node $n$ by \eqref{eq12}, where $\sigma$ is the elu activation function and GraphNorm is graph normalization \cite{Cai2021GraphNormAP}.
\begin{equation}
h_{n}^{l} = \sigma\left( GraphNorm ( W_{u1}^{l}v_{n}^{l} + W_{u2}^{l}h_{n}^{l-1} + b_{u}^{l} )  \right)\label{eq12}
\end{equation}

The $W$ above are all the weight matrix, and the $b$ above are all bias vectors, which will be learned by training. All of the attention mechanisms mentioned above use multiple heads.

\subsection{Classification}
When we obtain the node hidden vectors $h_{1}^{L},h_{2}^{L},\ldots,h_{|N|}^{L}$ from the last layer, we utilize attention mechanism to aggregate the information of each node to obtain vector representation $r$ as \eqref{eq13}\eqref{eq14}, where $g \in \mathbb{R}^{C_2}$ is a learnable vector.
\begin{equation}
\alpha_{n} = Softmax\left( {g^{T}h_{n}^{L}} \right)\label{eq13}
\end{equation}
\begin{equation}
r = {\sum_{n \in N}{\alpha_{n}h_{n}^{L}}}\label{eq14}
\end{equation}

To obtain the final classification prediction, we use an MLP, which is expressed as \eqref{eq15}.
\begin{equation}
pred = Softmax\left( MLP(r) \right)\label{eq15}
\end{equation}
The attention mechanism above is also multi-head attention. We employ the standard cross-entropy loss function for the training.

\section{Evaluation}
We implement code by torch\_geometric\footnote{https://pytorch-geometric.readthedocs.io/en/latest/index.html}. Our implementation is available on \href{https://github.com/qiankunmu/HDHGN}{\textcolor{black}{https://github.com/qiankunmu/HDHGN}}.

\subsection{Datasets}
We use Python800 and Java250 to train and assess our model. The two public datasets are from Project CodeNet \cite{puri2021codenet} which are obtained from downloading submissions from two online judge websites: AIZU Online Judge and AtCoder. The code snippets are classified by the problem. The statistics of the datasets are depicted in Table~\ref{t1}. To be clear, the AST node type means the AST type such as ``Module" and ``Assign," different from node types in HDHG, i.e., ``AST" and ``identifier." The edge type means the field name or called attribute in AST. We randomly split the dataset into the training set, validation set, and test set by 6:2:2.
\begin{table}[b]
\centering
\caption{Statistics of the datasets}
\label{t1}
\begin{tabular}{c c c}
\toprule[1pt]
 &Python800 & Java250 \\
\midrule
 Size & 240000 & 75000\\
 Labels & 800 & 250 \\
 Avg. node & 202.05 & 216.43 \\
 Avg. edge & 187.25 & 198.51 \\
 AST node types & 93 & 58\\
 Edge types & 58 & 61 \\
 Language & Python & Java \\
\bottomrule
\end{tabular}
\end{table}

\subsection{Baselines}
We compare our model with AST-based and GNN-based techniques which acquire the best performance in code classification including TBCNN \cite{Mou2016ConvolutionalNN}, TreeCaps \cite{Bui2021TreeCapsTC}, GGNN \cite{allamanis18learning}, GREAT \cite{Hellendoorn2020GlobalRM} and HPG+HGT \cite{Wang2022LearningTR}. TBCNN used a tree-based convolutional neural network to extract features from AST. A model called TreeCaps combines TBCNN and capsule networks. By adding edges like control flow and data flow to AST, the GGNN processes graphs from source code. GREAT is a model extracting global relational information from code graphs based on the transformer architecture. A technique known as HPG+HGT uses a heterogeneous graph transformer to describe code as a heterogeneous graph. We also trained a GCN \cite{Kipf2017SemiSupervisedCW} and a GIN \cite{Xu2019HPGNN} in an experiment to compare.

\subsection{Experiment Settings}
We use a parser from the official Python 3.8 ast library and javalang library\footnote{https://pypi.org/project/javalang/} to parse the code snippets into ASTs. The embedding vectors are produced by random initialization and learned via training. Our model's layer number was set to four. The hidden vector dimension size and embedding vector dimension size were both set to 128. We use a multi-head attention \cite{Vaswani2017AAN} mechanism and set the number of heads to eight. We employed Adam optimizer with the learning rate of $5 \times 10^{-5}$ to train our model. We set the dropout rate to 0.2. We optimized the hyper-parameters of other baselines for the validation set's greatest performance. The models were trained for 100 epochs and we saved the models which perform best in validation set. 

\subsection{Results}
We use the performance of the model on the test set as the outcome. We select classification accuracy as the metric. We calculate the mean accuracy and standard deviation after five times of the experiment. The results are depicted in Table~\ref{t2}. Our HDHGN outperforms other methods in both datasets. In Python800, our HDHGN is 2.88\% higher than the best baseline. In Java250, our model outperforms baseline models by at least 2.47\%. This demonstrates that our model utilizes the semantic and structural features of code AST more effectively than previous approaches.

\begin{table}[t]
\centering
\caption{Accuracy of models in code classification (in \%).}
\label{t2}
\begin{tabular}{c c c c c c}
\toprule[1pt]
& \multirow{2}{*}{Models} & \multicolumn{2}{c}{Python800} & \multicolumn{2}{c}{Java250} \\
& & mean & sd & mean & sd \\
\midrule
\multirow{2}{*}{AST-based} & TBCNN & 91.16 & $\pm0.10$ & 90.47 & $\pm0.10$ \\
&TreeCaps & 90.33 & $\pm0.11$ & 91.38 & $\pm0.11$ \\
\midrule
\multirow{5}{*}{GNN-based} & GCN & 91.94 & $\pm0.10$ & 90.01 & $\pm0.10$ \\
& GIN & 93.23 & $\pm0.10$ & 90.72 & $\pm0.10$ \\
&GGNN & 90.13 & $\pm0.11$ & 89.97 & $\pm0.11$ \\
&GREAT & 93.33 & $\pm0.12$ & 93.17 & $\pm0.11$ \\
&HPG+HGT\tablefootnote{ We use the outcomes that were reported in their research because the paper did not make their code publicly available. } & 94.99 & - & 93.95 & - \\
\midrule
Ours &HDHGN & \textbf{97.87} & $\pm0.10$ & \textbf{96.42} & $\pm0.10$ \\
\bottomrule
\end{tabular}
\end{table}

\begin{table}[t]
\centering
\caption{Accuracy of ablation study on Python800 (in \%).}
\label{t3}
\begin{tabular}{l c c}
\toprule[1pt]
 Variant & mean & sd \\
\midrule
HDHGN & \textbf{97.87} & $\pm0.10$ \\
 - hyperedge & 94.79 & $\pm0.11$ \\
 - heterogeneous information & 95.23 & $\pm0.11$\\
 - direction & 95.49 & $\pm0.10$\\
\bottomrule
\end{tabular}
\end{table}

\subsection{Ablation Study}
We perform some ablation studies of our HDHGN on Python800. We take into account three variants as below.
\subsubsection{- hyperedge}
We eliminate hyperedges from our model, leaving only paired edges, or normal edges, in the graph. A few regular edges will develop from the initial hyperedge.
\subsubsection{- heterogeneous information}
We eliminate heterogeneous information from our model, which entails treating identifier nodes and AST nodes as a single type of node in the graph and eliminating the information about edge types.
\subsubsection{- direction}
We remove direction information in our model, this means that the hyperedge is not directed hyperedge, it does not differentiate the head nodes and tail nodes.

We also repeat the experiment five times and compute the mean accuracy and standard deviation. The outcomes are depicted in Table~\ref{t3}. Removing hyperedge make the result decrease by 3.08\%. This demonstrates that high-order data correlations between AST nodes in code are indeed useful for comprehending programs. The removal of heterogeneous information reduces the result by 2.64\%. Heterogeneous information often contains a lot of semantic information, which is helpful for program understanding. Removing direction caused a drop of 2.38\% on the result. The direction of the graph can help enhance the model and get structural information by indicating whether the nodes connected by hyperedges are parent nodes or child nodes. The above outcomes demonstrate that our model can obtain a better understanding of AST structure and acquire more precise results in code classification after considering high-order data correlations, heterogeneous information, and direction information.

\section{Conclusion}
In this study, we propose an HDHGN for code classification. To possibly encode high-order data correlations between nodes of the same field or called attribute in AST, we introduce the use of hypergraphs. Due to the general hypergraph being homogeneous and undirected which will result in a lack of semantic and structural information, we propose an HDHG to represent AST. We propose an HDHGN accordingly to utilize high-order data correlation, heterogeneous information and direction information better than previous methods. We test our model using open Python and Java datasets, and we compare the results to the SOTA baselines developed using the AST and GNN. The experiment demonstrates that our HDHGN outperforms the baselines. Further ablation study describes that the HDHGN enhances the performance of code classification.

Presently, the hypergraph we produce is large and contains many nodes and edges. Future research will focus on ways to scale down hypergraphs for modeling AST and enhance the current hypergraph model to make it more effective at classifying codes.

\section*{Acknowledgment}
This work was supported by the Open Research Fund of NPPA Key Laboratory of Publishing Integration Development, ECNUP. 

\bibliographystyle{ieeetr}
\bibliography{ref}

\begin{thebibliography}{10}

\bibitem{Mou2016ConvolutionalNN}
L.~Mou, G.~Li, L.~Zhang, T.~Wang, and Z.~Jin, ``Convolutional neural networks
  over tree structures for programming language processing,'' in {\em AAAI},
  2016.

\bibitem{Gilda2017SourceCC}
S.~Gilda, ``Source code classification using neural networks,'' {\em 2017 14th
  International Joint Conference on Computer Science and Software Engineering
  (JCSSE)}, pp.~1--6, 2017.

\bibitem{Vagavolu2021AMO}
D.~Vagavolu, K.~C. Swarna, and S.~Chimalakonda, ``A mocktail of source code
  representations,'' in {\em ASE}, 2021.

\bibitem{Wang2022LearningTR}
W.~Wang, K.~Zhang, G.~Li, and Z.~Jin, ``Learning to represent programs with
  heterogeneous graphs,'' {\em 2022 IEEE/ACM 30th International Conference on
  Program Comprehension (ICPC)}, pp.~378--389, 2022.

\bibitem{Zhang2019ANN}
J.~Zhang, X.~Wang, H.~Zhang, H.~Sun, K.~Wang, and X.~Liu, ``A novel neural
  source code representation based on abstract syntax tree,'' {\em 2019
  IEEE/ACM 41st International Conference on Software Engineering (ICSE)},
  pp.~783--794, 2019.

\bibitem{Lu2021StudentPC}
M.~Lu, Y.~Wang, D.~Tan, and L.~Zhao, ``Student program classification using
  gated graph attention neural network,'' {\em IEEE Access}, vol.~9,
  pp.~87857--87868, 2021.

\bibitem{alon2018codeseq}
U.~Alon, S.~Brody, O.~Levy, and E.~Yahav, ``code2seq: Generating sequences from
  structured representations of code,'' in {\em ICLR}, 2019.

\bibitem{Alon2019code2vecLD}
U.~Alon, M.~Zilberstein, O.~Levy, and E.~Yahav, ``code2vec: learning
  distributed representations of code,'' {\em Proceedings of the ACM on
  Programming Languages}, vol.~3, pp.~1 -- 29, 2019.

\bibitem{Hu2018DeepCC}
X.~Hu, G.~Li, X.~Xia, D.~Lo, and Z.~Jin, ``Deep code comment generation,'' {\em
  2018 IEEE/ACM 26th International Conference on Program Comprehension (ICPC)},
  pp.~200--20010, 2018.

\bibitem{Liu2021RetrievalAugmentedGF}
S.~Liu, Y.~Chen, X.~Xie, J.~Siow, and Y.~Liu, ``Retrieval-augmented generation
  for code summarization via hybrid gnn,'' in {\em ICLR}, 2021.

\bibitem{Jiang2007DECKARDSA}
L.~Jiang, G.~Misherghi, Z.~Su, and S.~Glondu, ``Deckard: Scalable and accurate
  tree-based detection of code clones,'' {\em 29th International Conference on
  Software Engineering (ICSE)}, pp.~96--105, 2007.

\bibitem{White2016DeepLC}
M.~White, M.~Tufano, C.~Vendome, and D.~Poshyvanyk, ``Deep learning code
  fragments for code clone detection,'' in {\em ASE}, 2016.

\bibitem{allamanis18learning}
M.~Allamanis, M.~Brockschmidt, and M.~Khademi, ``Learning to represent programs
  with graphs,'' in {\em ICLR}, 2018.

\bibitem{Long2022MultiViewGR}
T.~Long, Y.~Xie, X.~Chen, W.~Zhang, Q.~Cao, and Y.~Yu, ``Multi-view graph
  representation for programming language processing: An investigation into
  algorithm detection,'' in {\em AAAI}, 2022.

\bibitem{Huang2021UniGNNAU}
J.~Huang and J.~Yang, ``Unignn: a unified framework for graph and hypergraph
  neural networks,'' in {\em IJCAI}, 2021.

\bibitem{Feng2018HypergraphNN}
Y.~Feng, H.~You, Z.~Zhang, R.~Ji, and Y.~Gao, ``Hypergraph neural networks,''
  in {\em AAAI}, 2018.

\bibitem{puri2021codenet}
R.~Puri, D.~S. Kung, G.~Janssen, W.~Zhang, G.~Domeniconi, V.~Zolotov, J.~Dolby,
  J.~Chen, M.~Choudhury, L.~Decker, V.~Thost, L.~Buratti, S.~Pujar, S.~Ramji,
  U.~Finkler, S.~Malaika, and F.~Reiss, ``Codenet: A large-scale ai for code
  dataset for learning a diversity of coding tasks,'' 2021.

\bibitem{Bui2021TreeCapsTC}
N.~D.~Q. Bui, Y.~Yu, and L.~Jiang, ``Treecaps: Tree-based capsule networks for
  source code processing,'' in {\em AAAI}, 2021.

\bibitem{Hellendoorn2020GlobalRM}
V.~J. Hellendoorn, C.~Sutton, R.~Singh, P.~Maniatis, and D.~Bieber, ``Global
  relational models of source code,'' in {\em ICLR}, 2020.

\bibitem{Berge1973GraphsAH}
C.~Berge, ``Graphs and hypergraphs,'' 1973.

\bibitem{Gallo1993DirectedHA}
G.~Gallo, G.~Longo, and S.~Pallottino, ``Directed hypergraphs and
  applications,'' {\em Discret. Appl. Math.}, vol.~42, pp.~177--201, 1993.

\bibitem{Sun2012MiningHI}
Y.~Sun and J.~Han, ``Mining heterogeneous information networks: Principles and
  methodologies,'' in {\em Mining Heterogeneous Information Networks:
  Principles and Methodologies}, 2012.

\bibitem{Vaswani2017AAN}
A.~Vaswani, N.~Shazeer, N.~Parmar, J.~Uszkoreit, L.~Jones, A.~N. Gomez,
  L.~Kaiser, and I.~Polosukhin, ``Attention is all you need,'' in {\em Advances
  in Neural Information Processing Systems 30: Annual Conference on Neural
  Information Processing Systems}, pp.~5998--6008, 2017.

\bibitem{Cai2021GraphNormAP}
T.~Cai, S.~Luo, K.~Xu, D.~He, T.-Y. Liu, and L.~Wang, ``Graphnorm: A principled
  approach to accelerating graph neural network training,'' in {\em ICML},
  2021.

\bibitem{Kipf2017SemiSupervisedCW}
T.~N. Kipf and M.~Welling, ``Semi-supervised classification with graph
  convolutional networks,'' in {\em ICLR}, 2017.

\bibitem{Xu2019HPGNN}
K.~Xu, W.~Hu, J.~Leskovec, and S.~Jegelka, ``How powerful are graph neural
  networks?,'' in {\em ICLR}, 2019.

\end{thebibliography}

\end{document}